\documentclass[aps,prl,reprint,groupedaddress,showpacs]{revtex4-1}
\usepackage{graphicx}
\usepackage[pdfstartview=FitH, pdfpagemode=UseNone]{hyperref}

\begin{document}


\title{Electronic Detection of Collective Modes of an Ultracold Plasma}


\author{K. A. Twedt and S. L. Rolston}
\affiliation{Joint Quantum Institute and Department of Physics, University of Maryland, College Park, Maryland 20742, USA}


\date{\today}

\begin{abstract}
Using a new technique to directly detect current induced on a nearby electrode, we measure plasma oscillations in ultracold plasmas, which are influenced by the inhomogeneous and time-varying density and changing neutrality. Electronic detection avoids heating and evaporation dynamics associated with previous measurements and allows us to test the importance of the plasma neutrality.  We apply dc and pulsed electric fields to control the electron loss rate and find that the charge imbalance of the plasma has a significant effect on the resonant frequency, in excellent agreement with recent predictions suggesting coupling to an edge mode.  
\end{abstract}


\maketitle

Collective oscillations are central to the study of plasmas as they embody the rich physics unique to the plasma state and provide diagnostics of plasma density and temperature.  Ultracold neutral plasmas (UCPs) \cite{Killian2007pr,Killian2010pt} are a novel system for the study of collective behavior as they have extremely low temperatures (1-100 K), an inhomogeneous density and are on the border of strong coupling.  Previous experiments on UCPs have observed plasma oscillations \cite{Kulin2000}, Tonks-Dattner resonances \cite{Fletcher2006}, ion acoustic waves \cite{Castro2010, McQuillen2011} and a high-frequency electron drift instability \cite{Zhang2008b}.  

Like most laser-produced plasmas, UCPs are freely expanding systems.  External fields can force varying rates of electron evaporation, which gives a time-dependence to the global neutrality, an important feature of these systems that has only recently been explored \cite{Twedt2010, Lyubonko2010}.  A recent theoretical study of plasma oscillations \cite{Lyubonko2010}, the simplest collective mode, considered the impact of electron loss and predicted the existence of a zero-temperature mode with a resonant frequency that increases for less neutral plasmas.  The theory assumed a spherically symmetric electron spatial distribution, $n_e$, but we have found a significant asymmetry in typical experiments \cite{Twedt2010} and the effect of this on electron oscillations has not been addressed. 

Here, we excite and detect plasma oscillations of an UCP taking into account the changing density, neutrality and symmetry of $n_e$.  We find the variation of the resonant frequency with neutrality agrees well with the predictions of \cite{Lyubonko2010}.  In addition, we have developed a new diagnostic for UCPs where we directly detect oscillations through the current induced on a nearby electrode.  The method is more accurate than previous measurements based on electron emission, can be used when charged particle detection is not possible, and gives UCP experiments access to a broader spectrum of collective modes.  

We create a plasma by two-photon ionization of about $10^6$ metastable Xe atoms collected in a magneto-optical trap \cite{killian1999}.  The initial plasma density is roughly Gaussian with an rms radius of 0.3 - 0.6 mm. The initial energy given to the electrons, $E_{e}$, is controlled by tuning the energy of the ionization laser above the ionization limit.  After creation, the plasma loses a few percent of the electrons until a sufficient charge imbalance exists to trap the remaining electrons, forming a plasma.  The plasma expands, driven by thermal electron pressure.  The ion distribution, $n_i$, remains  Gaussian, following a self-similar expansion \cite{laha2007} described by $\sigma_{i}^2 (t) = \sigma_{i}^2 (0) + v_{0}^2 t^2$, where $\sigma_{i}$ is the rms radius of the ion distribution and the expansion velocity is typically $v_{0}=50-100$ m/s, determined by $E_{e}$.  

\begin{figure}[b]
\includegraphics[width=0.48\textwidth, viewport = 91 240 517 550]{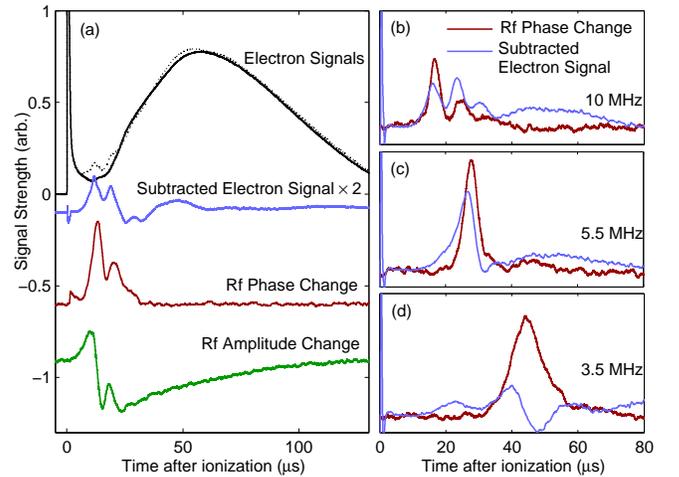}
\caption{Comparison of rf signals to electron signals. (a) The dotted and solid black lines are the electron emission signals with and without a $f = \omega_{\mathrm{rf}}/2\pi = 14$ MHz rf field applied.  The rf phase and amplitude signals are the difference of the signal obtained with and without the plasma, have been rescaled by the same arbitrary factor and are offset for clarity.  (b)-(d) Rf phase change compared to subtracted electron signal for three different frequencies. \label{RfSignalCompare}}
\end{figure}

Two wire mesh grids located 1.4 cm on either side of the plasma apply a weak electric field (5-10 mV/cm) that directs evaporating electrons out of the plasma region and onto a microchannel plate detector.  A typical electron current signal including the prompt loss of electrons and the electron evaporation during expansion is shown in Fig.~\ref{RfSignalCompare}(a).   

UCPs are small systems, consisting of only $10^{4} - 10^{9}$ ions and electrons and sizes of 0.1 mm to 1 cm, so available experimental probes have been limited.  Optical absorption and fluorescence imaging of the plasma ions \cite{simien2004, cummings2005} have provided spatially resolved density and velocity measurements of the ions.  Information about the electrons has predominantly been obtained by monitoring the loss rate on a charged particle detector as described.  This method of electron detection has succeeded in observing plasma oscillations \cite{Kulin2000,Fletcher2006}, but only indirectly by applying a constant driving field that resonantly heats the electrons and observing an enhanced loss rate.  Thus the measurements are subject to the dynamics involved with heating the electrons and their subsequent evaporation.  

We present a new approach to studying electron resonances in UCPs by directly measuring changes in the rf field.  Measurements of rf absorption are commonplace in low density laboratory plasmas.  Most analogous to our system are measurements of zero-temperature oscillations in non-neutral plasmas trapped in Penning traps \cite{Wineland1975, Weimer1994, Tinkle1995, ATHENA2003pop}.  These plasmas are typically of similar size and density to our neutral plasmas, but our rapidly expanding plasmas are untrapped so the measurements must be made during the fast time evolution of the plasma density.  Our resonances last only a few $\mu$s, about 100 times shorter than the averaging times used for non-neutral plasmas.  

\begin{figure}
\includegraphics[width=0.48\textwidth, viewport = 0 131 230 207]{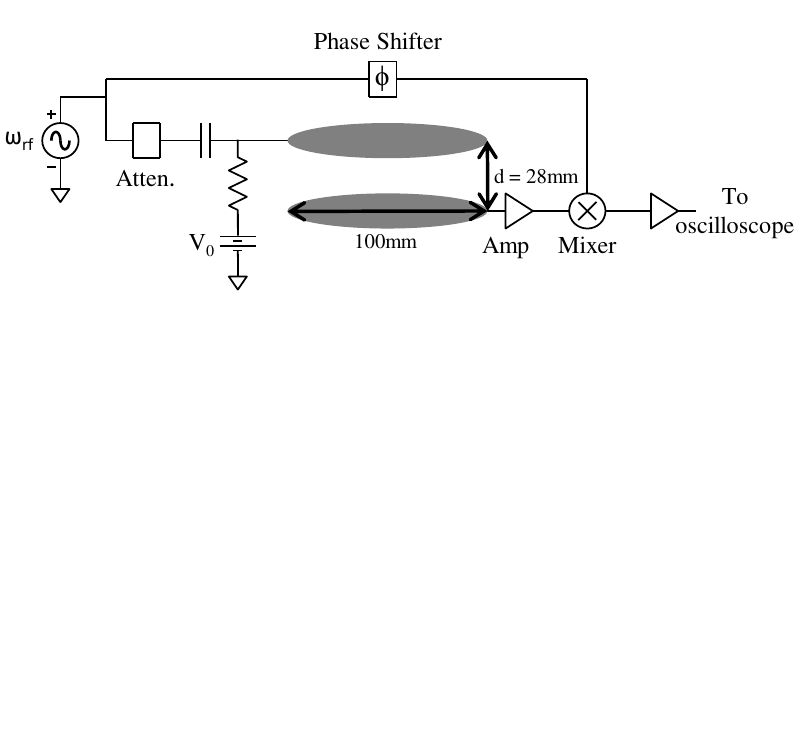}
\caption{Schematic for homodyne detection of plasma oscillations.  The rf source is attenuated (Atten.) and applied to the top grid. The dc bias and voltage pulses are applied at $V_{0}$.  \label{CircuitSchematic}}
\end{figure}

We detect plasma modes  by applying a weak, continuous rf drive at frequency $\omega_{\mathrm{rf}}$ to the grid located above the plasma and monitor the amplitude and phase changes of the voltage coupled to the symmetric grid below, $V_{b}$, as sketched in Fig.~\ref{CircuitSchematic}.  In the absence of a plasma, $V_{b}$ is constant in time and simply related to the electrode geometry.  When a plasma is present and driven near a resonant frequency, the oscillation of the plasma induces a current on the bottom grid that interferes with the background signal.  The signal from the plasma is much smaller than the background, so we observe only small changes in the amplitude and phase of $V_{b}$ as the plasma density quickly scans through resonance with the driving field.  The results of both quadratures of this homodyne measurement are shown in Fig.~\ref{RfSignalCompare}(a).  All measurements are done with $\geq 400$ kHz bandwidth, sufficiently large to capture the fast changes in the rf signals.  Due to this large bandwidth and small particle number, the signal-to-noise ratio on a single experimental shot is often less than 1, so all plots are an average of at least 150 shots.  

We qualitatively reproduce the shapes of the rf signals using a circuit model that treats the plasma as a series RLC oscillator in close analogy with \cite{Wineland1975}.  This shows that a peak in the phase-change signal corresponds to the time at which the plasma is resonant with the applied field, and thus we will focus on this signal in the remainder of this paper.  A more detailed analysis could also provide information about mode damping and the electron temperature.  

Figure~\ref{RfSignalCompare} shows the comparison of the rf measurements with the previously employed enhanced electron emission method.  The earliest-time peak in both the rf and electron signals is the zero-temperature plasma resonance.  For large plasmas and higher $\omega_{\mathrm{rf}}$, we routinely see extra peaks at later time in the electron signal, e.g. three peaks in parts (a) and (b) of Fig.~\ref{RfSignalCompare}, that were previously identified as temperature-dependent Tonks-Dattner resonances \cite{Fletcher2006}.  We have difficulty resolving more than one extra peak in the rf signals.  For larger driving amplitude we can resolve a total of three peaks in the rf signals, but the third is a factor of 10 or more smaller in amplitude than the first.  We anticipate that a higher signal-to-noise measurement would allow us to see more of the modes.  Molecular dynamics simulations \cite{LyubonkoComm} have also found multiple peaks in the rf absorption in UCPs, but a detailed understanding of these modes is still lacking.  We focus our analysis on the zero-temperature resonance. 

The resonance times in the two methods agree to within 1 $\mu$s in most cases, after correcting for a delay due to finite bandwidth.  There is disagreement at low $\omega_{\mathrm{rf}}$ ($<6$ MHz), where the electron signal becomes less reliable.  The electron signal measures evaporation caused by rf heating, but we should not expect a linear relationship between energy absorption and electron emission.  As the heating begins and electrons are lost, the plasma potential well deepens, so a greater input energy is needed to subsequently maintain the same electron flux.  This effect is most evident at low frequencies as the plasma response becomes much broader in time.  By contrast, the rf signals measure an induced current that is directly proportional to the amplitude of electron oscillations.

\begin{figure}
\includegraphics[width=0.4\textwidth, viewport = 46 35 357 244]{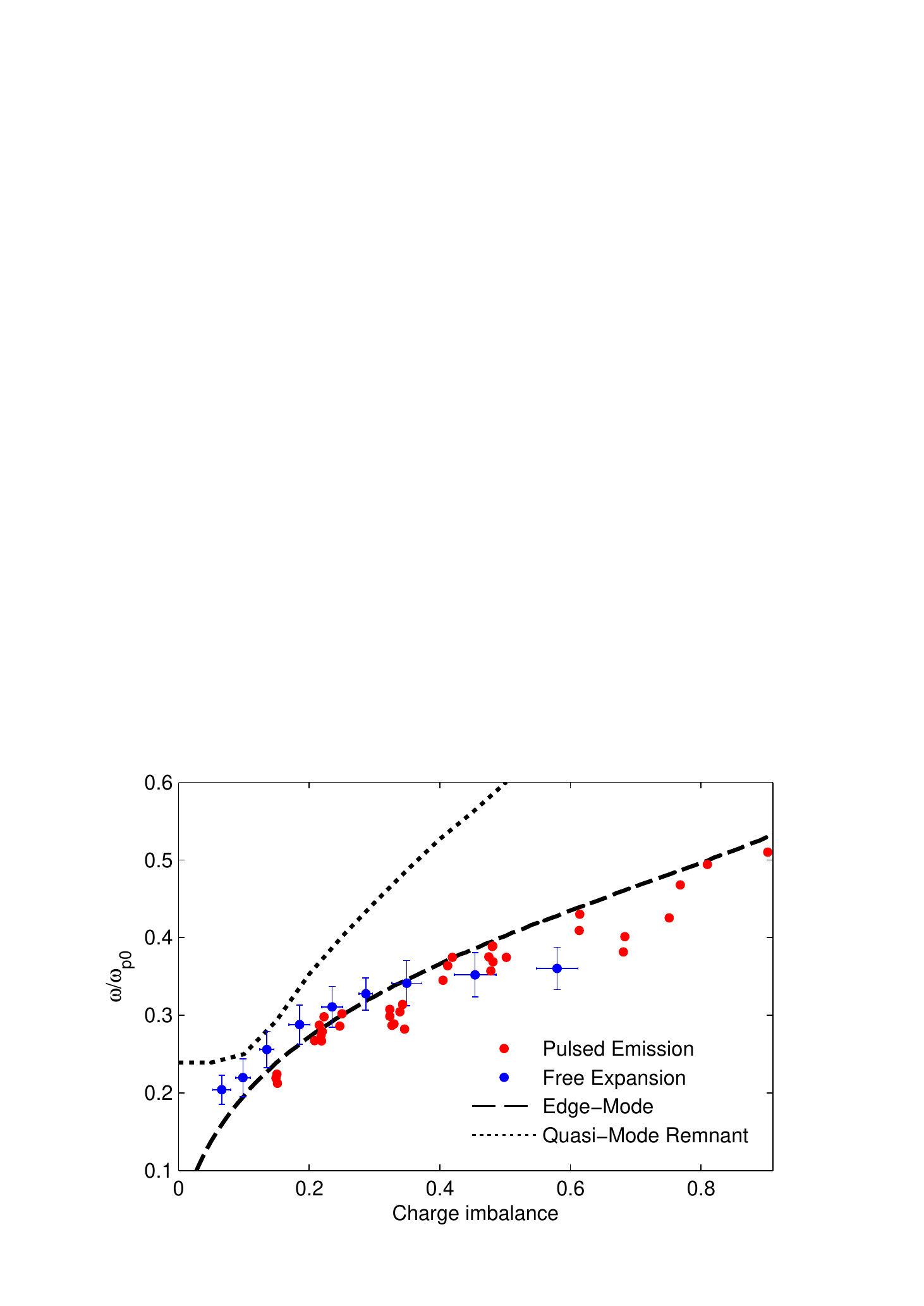}
\caption{Edge-mode theory compared to free expansion and pulsed emission data.  The edge-mode and quasi-mode remnant curves are from \cite{Lyubonko2010}.  Blue points with error bars are the average of many free expansion measurements.  Red points are individual measurements after pulsed electron emission.  Results are the combination of many experimental runs all with $E_e/k_b = 100$ K and $6\times10^5 < N_i < 10^6$.  \label{EdgeModeData}}
\end{figure}

To understand the cold plasma resonance, we consider the full picture of the spatial distribution of electrons.  Optical measurements of the plasma ions have shown that $n_i$ remains Gaussian throughout  expansion \cite{laha2007}.  At the center of the plasma, $n_e$ must be nearly equal to $n_i$, but electron loss ensures deviations at the plasma edge.  Bergeson and Spencer \cite{Bergeson2003} solved the cold plasma fluid equations for a perfectly neutral plasma ($n_e = n_i$ in all space), assuming no electron loss, and found only a single quasi-mode with a maximum energy absorption at a frequency $\omega = 0.24\omega_{p0}$, where $\omega_{p0} = \sqrt{e^{2}n_{e0}/m_{e}\epsilon_{0}}$ and $n_{e0}$ is the central plasma density.  But even early in the plasma lifetime the deviation from neutrality may be non-negligible, owing to the prompt loss of electrons at plasma creation [Fig.~\ref{RfSignalCompare}(a)].  Lyubonko \textit{et al.} \cite{Lyubonko2010} allowed for electron loss by treating $n_e$ as a truncated Gaussian with the truncation radius set by the charge imbalance $\delta = (N_{i}-N_{e})/N_{i}$, where $N_{i}$ and $N_{e}$ are the number of ions and electrons.  For plasmas with any significant charge imbalance ($\delta \geq 5\%$), they found a mode where the majority of energy was absorbed near the sharp edge of the cold electron distribution.  The relative frequency of the edge-mode resonance, $\omega_{rel} = \omega / \omega_{p0}$, increases with $\delta$ as shown in Fig.~\ref{EdgeModeData}.  

\begin{figure}
\includegraphics[width=0.38\textwidth, viewport = 170 300 425 485]{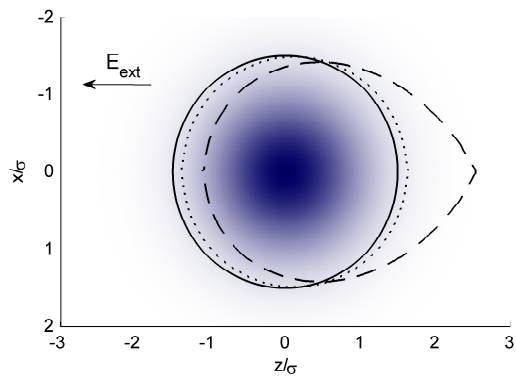}
\caption{Electron distributions with $\delta=0.53$.  The shaded region represents the Gaussian ion distribution.  Cold electron distributions match $n_i$ inside the boundary lines and drop sharply to zero outside.  The solid line is the symmetric electron boundary from theory.  The dashed line is the electron boundary during free expansion and the dotted line is the boundary after a pulsed emission of electrons, both of which have an asymmetry caused by the external field, $E_{ext}$.    \label{EdistCompare}}
\end{figure}

We perform two types of experiments to test the edge-mode prediction.  First, we record the resonance times from the rf phase change signal during normal free expansion of the plasma.  Figure~\ref{RfSignalCompare} shows typical measurements for different $\omega_{\mathrm{rf}}$.  For smaller $\omega_{\mathrm{rf}}$, the resonances are observed later in time, corresponding to a lower density.  But due to the continuous electron loss, later time also corresponds to larger charge imbalance, which should increase the relative resonant frequency.  We fit a Gaussian to the first peak of the rf phase change to get the resonance time.  We independently measure the plasma expansion velocity and ion number, which allows us to calculate the relative resonant frequency $\omega_{rel}$.  The charge imbalance is calculated by integrating the electron signals.  Only a small 10-15 mV/cm electric field is needed to collect all plasma electrons on our detector.  The fraction of the integrated current that arrives before the resonance time gives the charge imbalance $\delta$.  The results are shown in Fig.~\ref{EdgeModeData}.  

A potential problem with this test is that we expect $n_e$ in the experiment to have a different shape than the symmetric truncated Gaussian used in theory.  We have found that $n_e$ in a freely expanding plasma can develop a strong asymmetry from the influence of externally applied or stray dc electric fields, and we intentionally apply such a field to facilitate electron detection.  This field perturbs $n_e$, which we have used to explain the observed rate of electron loss from our system \cite{Twedt2010}.  The difference between our calculated $n_e$ and the symmetric case for the same charge imbalance is shown in Fig.~\ref{EdistCompare}.  

\begin{figure}
\includegraphics[width=0.48\textwidth, viewport = 100 315 518 474]{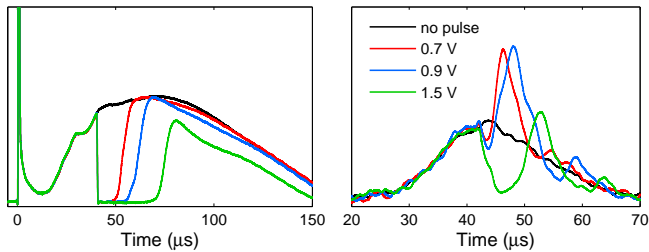}
\caption{Electron signals (left) and rf phase change signals (right) with a voltage pulse of 1 $\mu$s length and varying amplitude applied at 40 $\mu$s.  The applied frequency is $f=4$ MHz.  \label{PulsedEmission}}
\end{figure}

In a second experiment, we dump electrons from the plasma with short voltage pulses of 0.5-2 $\mu$s, chosen to be longer than the electron collision time.  This creates an $n_e$ (dotted line in Fig.~\ref{EdistCompare}) that is closer to the symmetric distribution used in theory and gives more control over $\delta$.  Fig.~\ref{PulsedEmission} shows examples of the electron emission signal with pulses applied.  

Before the voltage pulse, we assume a Gaussian $n_i$ and an asymmetric $n_e$, like the dashed line in Fig.~\ref{EdistCompare}.  After the pulse, the remaining electrons will be concentrated mostly at the center of $n_i$.  The dc electric field will still polarize the plasma, but the electrons are now held in a deep well and there are not enough to reach the edge of the ions.  Electron emission ceases for 10 to 20 $\mu$s.  As the plasma continues to expand, the well depth decreases and $n_e$ slowly becomes less symmetric until it again reaches the edge of $n_i$ and electron emission returns.  We can calculate $n_e$ after a pulse using the same algorithm used in \cite{Twedt2010} but fixing the value of $\delta$.  Fig.~\ref{EdistCompare} shows one example result.  

The drop in $N_e$ must affect the ion expansion, as the ions outside of the electron cloud are free to move in response to both the external field and their own interactions without electron screening.  For larger electron dumps, the density of free ions is large enough that a significant Coulomb explosion occurs and disrupts the following neutral plasma expansion.  The electron signal following a pulsed emission is an indicator of the effect of this Coulomb explosion on the plasma.  Since the electron loss rate is related to the size and shape of the ion cloud, an electron signal very similar to that during normal free expansion indicates a minimal change in plasma expansion.  As shown in Fig.~\ref{PulsedEmission}, we can for lower voltage pulses get excellent agreement in the late-time electron signals.  

For a given voltage pulse, we adjust $\omega_{\mathrm{rf}}$ and look for resonances in the rf phase change during the dead time of electron emission, as shown in Fig.~\ref{PulsedEmission}.  When we are near a resonance, a clear peak is observed  that we identify as the edge-mode resonance.  Changing the strength of the voltage pulse dumps more electrons, increasing $\delta$, but leaves all other parameters unchanged.  We see that this direct change in $\delta$ increases the time of resonance, which corresponds to an increase in $\omega_{rel}$ as predicted.  

The data in Fig.~\ref{EdgeModeData} are created by applying voltage pulses at many times during expansion.  At each time, we sweep $\omega_{\mathrm{rf}}$ to find resonance peaks that come just after the pulsed emission.  We calculate $\omega_{rel}$ from the resonance time assuming an uninterrupted ion expansion, and we only plot points where the observed peak comes within 7 $\mu$s of the end of the voltage pulse.  To get to higher $\delta$ we wait until later time in the expansion, letting the plasma naturally lose more charge, before applying the pulse.  During pulsed emission, we pulse the electrons away from the detector but still collect all other electrons.  We find $\delta$ by comparing the integrated current after the time of resonance to the full integrated current in the free expansion experiment.

It is clear from Fig.~\ref{EdgeModeData} that the free expansion and pulsed experiments agree with each other and the edge-mode theory.  Agreement in the free expansion data is worst at large $\delta$, but these rf signals are very broad, spanning 10s of $\mu$s.  Significant changes to collision rates and significant electron loss during the response time may affect the magnitude of the rf signal and distort the results.  The pulsed emission experiments give us control over $\delta$, and we can more easily probe large values. It is unclear why the rf signal after a pulsed emission has a stronger and sharper response than during free expansion.

The importance of the shape of $n_e$ is also not immediately clear.  During free expansion, we expect $n_e$ to be more asymmetric than the distribution after a pulsed emission, yet both measurements seem to agree equally well with the perfectly symmetric theory, suggesting that the spatial distribution of charge is only of secondary importance to the integrated amount of charge.  It seems surprising that the position of the electron cloud edge is not a larger factor given the finding in \cite{Lyubonko2010} that the large majority of energy is absorbed at this edge.  It would be interesting to see how the solution of the cold plasma fluid equations changes for the asymmetric distributions of Fig.~\ref{EdistCompare}.     

In conclusion, we have shown detailed measurements of cold plasma oscillations in an expanding ultracold plasma with a time-varying $n_e$.  Both of our experimental approaches support an increase of resonance frequency with charge imbalance, in agreement with a zero-temperature theory.  We have also presented a new diagnostic tool for probing oscillations in ultracold plasmas that is more accurate and more versatile than the previous method.  

An important feature of our new measurement is that it allows us to probe electron properties through their resonant behavior without the need for charged particle detection.  This advantage was evident in our ability to observe resonances during the dead time in electron emission.  Charged particle detection is also prevented when a magnetic field is applied transverse to the axis of the electric grids.  We have observed upper hybrid resonances in this setup, which will be the subject of future work.   

We have, in some regimes, also been able to observe the free decay of electron oscillations after abruptly turning off the rf drive.  Studying mode damping should provide information on the collision properties and electron temperature in UCPs.  It is worth noting that we observe plasma resonances at much later times than are normally studied in UCPs.  At 90 $\mu$s, the plasma density has dropped to $2 \times 10^5 \ \mathrm{cm}^{-3}$, but we can still observe clear collective behavior, which is a testament to the extremely low electron temperature, expected to be less than 1 K at that time.    
 
We thank A. Lyubonko, T. Pohl and J. M. Rost for helpful discussions and R. A. Perrotta for technical assistance.  This work was supported by the NSF PHY1004242.


\bibliography{rf}

\end{document}